\documentclass[aps,pre,onecolumn,amsmath,amssymb,10pt]{revtex4-1}

\usepackage{amssymb,amsfonts,amsmath,graphicx}
\usepackage{color}
\usepackage{natbib}
\usepackage{float}
\usepackage{bm}
\usepackage[colorlinks=true,citecolor=blue,urlcolor=blue]{hyperref}
\usepackage{soul}
\usepackage[utf8]{inputenc}

\def\be{\begin{equation}}
\def\ee{\end{equation}}
\def\bea{\begin{eqnarray}}
\def\eea{\end{eqnarray}}
\def\p{\partial} 
\def\pt{\partial_t} 
\def\nn{\nonumber}
\def\f{\frac}
\def\a{\alpha}
\def\b{\beta}
\def\d{\nabla}

\def\l{\left(}
\def\r{\right)}
\def\[{\left[}
\def\]{\right]}

\def\bs{\boldsymbol}

\def\cob{\color{blue}}

\newcommand{\bfr}{{\bf r}}

\newcommand{\bff}{{\bf f}}
\newcommand{\bfv}{{\bf v}}

\newcommand{\bfm}{{\bf m}}
\newcommand{\bfn}{{\bf n}}

\newcommand{\bfJ}{{\bf J}}

\usepackage[%Uncomment any one of the following lines to test 
margin=0.7in,
]{geometry}
\begin{document}

\title{Stratification, multivalency and turnover of the active cortical machinery are required for steady active contractile flows at the cell surface
}
% Force line breaks with \\
% \thanks{A footnote to the article title}%

\author{Sk Raj Hossein}
\affiliation{%
Raman Research Institute, Bangalore 560080, India
}%

\author{Rituparno Mandal\footnote{Previously: Simons Centre for the Study of Living Machines, National Centre for Biological Sciences (TIFR), Bellary Road, Bangalore 560065, India}}
\affiliation{%
Institute for Theoretical Physics, Georg-August-Universit\"at G\"ottingen, 37077 G\"ottingen, Germany
}%

\author{Madan Rao}
\email{madan@ncbs.res.in}
\affiliation{
Simons Centre for the Study of Living Machines, National Centre for Biological Sciences (TIFR), Bellary Road, Bangalore 560065, India
}

\begin{abstract}
Many cell membrane proteins that bind to actin form dynamic clusters
driven by contractile 
flows generated by the actomyosin machinery at the cell cortex. Recent evidence suggests that 
a necessary condition for the generation of these protein clusters on the membrane is the
stratified organization of the active agents - formin-nucleated actin, myosin-II minifilaments, and ARP2/3-nucleated actin mesh - within the cortex. 
Further, the observation that these clusters dynamically remodel,  requires that the components of this active machinery undergo turnover. 
Here we develop a coarse-grained agent-based Brownian dynamics simulation that incorporates the effects of stratification, binding of myosin minifilaments to multiple actin filaments and their turnover. We show that these three features of the active cortical machinery - {\it stratification}, {\it multivalency} and {\it turnover} - are critical for the realisation of a nonequilibrium steady state characterised by contractile flows and dynamic orientational patterning.
We show that this nonequilibrium steady state enabled by the above features of the cortex, can facilitate multi-particle encounters of membrane proteins that
 profoundly influence the kinetics of bimolecular reactions at the cell surface.
\end{abstract}

\pacs{47.63.mh, 87.17.-d, 05.60.cd}

\maketitle
\section{Introduction}
\label{sect:Intro}
Several studies show that the spatiotemporal 
organisation of cell membrane molecules at different scales is controlled by its interaction with
a thin layer of actomyosin cortex
that adjoins the cell membrane~\cite{rao2014active,goswami2008nanoclusters,gowrishankar2012active,saha2015diffusion,plowman2005h,lingwood2010lipid}.  Thus membrane proteins that bind to cortical actin are driven by actomyosin contractility to form dynamic nanoclusters.
Such studies have lead to the description of the cell surface as an {\it Active Composite} -- an asymmetric multicomponent bilayer membrane juxtaposed with an active
actomyosin layer~\cite{rao2014active,gowrishankar2012active,saha2015diffusion,chaudhuri2011spatiotemporal} -- wherein membrane
molecules that bind to cortical actin, are driven by contractile flows generated by active actomyosin stresses to form dynamic clusters.

The actomyosin cortex is primarily built from two ingredients, the actin cytoskeleton and Myosin motors.
The actin cytoskeleton is 
composed of dynamic linear polar filaments, nucleated by Formin~\cite{gowrishankar2012active,pruyne,sagot} together with an extensively branched, relatively static meshwork
nucleated by Arp2/3~\cite{rouiller,goley,morone2006three}. The major component of Myosin motors, is nonmuscle myosin-II, which assemble as large minifilaments consisting of $\sim\,30-50$ myosin heads~\cite{billington,pollard}. If myosin-II minifilaments are to drive the observed active nanoclustering of membrane proteins~\cite{rao2014active,goswami2008nanoclusters,gowrishankar2012active}, then one might worry
whether steric constraints imposed by these bulky structures could frustrate clustering?
A recent study, involving agent-based simulations and {\it in vitro} reconstitution experiments, showed that {\it stratification} of the components of the cortical machinery, with myosin-II
layered atop a layer of dynamic actin which in turn adjoins the membrane, resolves this potential conflict~\cite{koster2016actomyosin,amitdas2020}. Stratification can
circumvent the steric frustration due to myosin-II minifilaments and can drive contractile flows that draw in the dynamic actin filaments together to form a 
 variety of orientational patterns including asters~\cite{amitdas2020}.

Moreover {\it in vivo}, the orientational patterning of actin filaments at the cortex and the resulting nanoclustering of membrane proteins, is {\it dynamic} and maintained in a nonequilibrium steady state by steady active contractile flows~\cite{rao2014active,goswami2008nanoclusters,gowrishankar2012active,guha2005,medeiros2006,reichl2008,
yang2012}. This is 
in contrast to most {\it in vitro} reconstitution systems, where the orientation patterns of the contractile actomyosin get jammed and consequently the protein clusters, once formed, disperse and diffuse away~\cite{koster2016actomyosin,linsmeier2016,murrell2012,vogel2013}.
In order to {\it maintain} a nonequilibrium steady state characterised by steady active contractile flows, dynamic orientational patterning and dynamic protein clustering, one 
needs, in addition to the stratification of the active machinery, a constant {\it turnover} of active components ~\cite{reichl2008,sonal2019,salbreux2012,fakhri2018}, that gives rise to a 
dynamic force patterning, 
as has been realised in some recent  {\it in vitro} studies that employ a continuous ATP regenerating system~\cite{koster2016actomyosin,sonal2019,fakhri2018}.

Here we bring together, in a rather simple coarse-grained agent based simulation, all the  ingredients necessary for observing the nonequilibrium steady state described above, viz., stratification, turnover and 
 multivalency of force generators. We do this without having to explicitly include a structural model for the molecular force generators that would have made it computationally challenging to monitor long time dynamics. Thus our coarse-grained simulation method 
  is of interest in and of itself, since while not  explicitly incorporating myosin minifilaments, it takes into account its effect on {\it currents}, {\it forces} and {\it torques}
applied to single and multiple actin filaments; indeed  we show that 
 inhibition of any one of the three ingredients, leads to a loss of the desired phenotype. 
The multivalency and turnover of active components give rise to an activation/inhibition of local active stresses, a feature that is critical for the attainment of the nonequilibrium steady state~\cite{murthy2005,guha2005,medeiros2006,haviv2008,salbreux2012,fakhri2018,sonal2019,keren2017}.

We show that this simple coarse-grained model, which incorporates the fluctuating active forces and torques in a stratified geometry, recapitulates the nonequilibrium steady states observed both  {\it in vivo} and in properly designed reconstitution experiments. By coupling this actomyosin dynamics to the dynamics of proteins on the two dimensional membrane, we find that  this automatically drives {\it multi-particle} encounters and dynamical clustering of membrane proteins. 
This provides the motivation to study the influence of active stresses on the dynamics of generic bimolecular chemical reactions, $A + B\,\,{\rightleftharpoons}\,\,C$ on the cell surface, an issue of profound biological consequence~\cite{chaudhuri2011spatiotemporal,kholodenko1,Iyengar1,haugh}. 

\section{Active dynamics of cell surface molecules driven by actin filaments and myosin minifilaments in a stratified cortex}
\label{sect:Simulation}

\begin{figure}[H]
\centering
\includegraphics[height=0.25\linewidth]{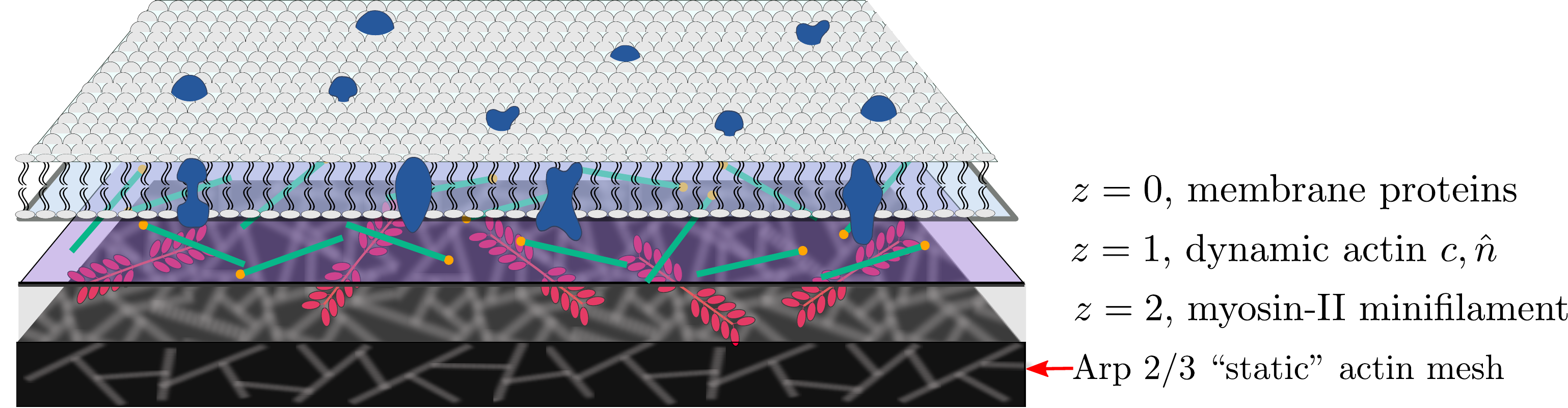}
\caption{Schematic of the cell surface as a stratified active composite of the multicomponent bilayer membrane ($z=0$), the thin layer of dynamic linear actin filaments ($z=1$) and 
a layer of myosin-II minifilaments ($z=2$) atop a dense Arp2/3 crosslinked actin mesh~\cite{kusumi2012}. The myosin-II minifilaments (red) bind to actin filaments (green) in $z=1$ and apply active forces and torques. In turn, the actin filaments bind to transmembrane proteins (blue) in $z=0$ and drive contractile flows leading to dynamic clustering of the proteins. The $+$ end of actin filaments is depicted by orange dots.}
\label{fig:schematic}
\end{figure}

Our coarse-grained simulation incorporates the three ingredients crucial to the active composite - stratification of the actomyosin cortex suggested by~\cite{amitdas2020}, multivalency of myosin-II minifilaments and turnover of the active components, and the
way they influence the dynamics of proteins that reside on the plasma membrane. 
To this end, we describe coarse-grained agents ({\cob{see}} Fig.\,\ref{fig:schematic}) along three different two-dimensional layers (strata) labelled $z=0$ (membrane), $z=1$ (dynamic formin-nucleated actin filament layer), and $z=2$ (myosin minifilament layer). The multivalent myosin minifilaments 
at $z=2$, stochastically bind and unbind  onto (possibly multiple) dynamic actin filaments at $z=1$; when bound they generate active 
forces and torques on the filaments. The dynamic actin filaments in turn drive contractile flows of the membrane proteins that bind to it.

Membrane proteins in the $z=0$ layer are modelled as mono-disperse soft discs of diameter $\sigma$. 
The dynamics of membrane proteins are determined by protein-protein interactions, interactions with the dynamic actin filaments situated in the 
$z=1$ layer and thermal noise. A pair of proteins separated by a distance $r_{\a\b}$ interact via a
purely repulsive potential, a truncated and shifted pair potential of the form,
\bea
V_{p}(r_{\a\b}) & = & 4 \epsilon \bigg(\frac{\sigma}{r_{\a\b}}\bigg)^{12} + V_0 +V_2 \;  r_{\a\b}^2 + V_4 \;  r_{\a\b}^4 \,\,\,\,\, \mbox{for} \,\, r_{\a\b}\leq \sigma  \nonumber \\
        & =  &   0    \,\,\,\,\, \mbox{for} \,\, r_{\a\b} > \sigma 
        \label{eq:partpot}
\eea
where 
the values of $V_0$, $V_2$ and $V_4$ are chosen so that the potential and force are continuous at the truncation point. In what follows, we 
set $\sigma=1$ and $\epsilon=1$ to be the units of length and energy, respectively (see Table 1 in Appendix).

Actin filaments in the $z=1$ layer are represented as a string of beads, the details of the inter-bead potential will be specified later in this section. Here, we focus on its influence
on the dynamics of a membrane protein in the $z=0$ layer, which is simply that when bound to a bead belonging to an actin filament, it moves with the velocity of that  filament, in a 
``no-slip'' manner. We implement this by prescribing the net force on the $\a$-th membrane protein in the $z=0$ layer as,

\be
    \bff_\a(t) =
    \begin{cases}
      \sum_{\b\neq \a} {\bf F}_{\a \b}^p(t) + {\bs \xi}_\a^p(t) & \text{(always)}\ \\
     -k\, \vert \bfr_{\alpha}-\bfr_{m,i}\vert & \text{(applicable when the protein is bound to any bead $m$ of filament $i$)}
     \end{cases}
    \label{eq:particle}
\ee

where ${\bf F}_{\a \b}^p$ is the force on the $\a$-th membrane protein due to the $\b$-th protein calculated from Eq.\,\ref{eq:partpot}, ${\bs \xi}_\a^p$ is the thermal noise, drawn from a Gaussian distribution with zero mean and variance 
$=2 k_B T \gamma_p/\Delta t$, where $\gamma_p$ is the friction coefficient of the proteins in the membrane,
and finally, $k$ is the attractive spring force between the the actin filament and the protein when it is  bound to it. The binding-unbinding status of each 
membrane protein is determined by switching rates $k_{b}$ and $k_{u}$, respectively, where binding to bead $m$ is initiated when a
distance criterion  $|\bfr_{\alpha}-\bfr_{m,i}| \leq 1.6\sigma$ is met. We vary $k_u$ over the range $10^{-5} - 10^{-1}$, where we use $\gamma_p$ to set the unit of time (see, Table\,1).
With this force, the position of the $\a$-th membrane protein gets updated by $\bfv_\a(t) = \bff_\a(t)/\gamma_p$ in a time $\Delta t$.

The dynamics of the actin filaments take place in the layer $z=1$.
Since the linear formin-nucleated actin filaments are much shorter than the persistence length $\ell_p$, they act as rigid rods of length $l$ and diameter $b$, with $l \gg b$. Our agent based update rules for the dynamics of actin filaments are motivated by the following - in the limit of {\it dilute}
concentration of actin filaments, we can 
define $c(\bfr, \hat \bfn, t)$ as the single filament distribution function of actin filaments of centre-of-mass position $\bfr =(x,y)$ and polar orientation $\hat \bfn=(\cos \theta, \sin \theta)$, which
obeys the Smoluchowski equation~\cite{liverpool2003,ahmadi2006},
\be
\pt c(\bfr, \hat \bfn, t) = - \d \cdot {\bf J} - \mathcal{R} \cdot  \mathcal{J}
\label{eq:smoluchowski}
\ee
where ${\bf \mathcal{R}} = \hat \bfn \times \p_{\hat \bfn}$ is the rotational operator.
The translational and rotational currents $\bfJ$ and $\mathcal{J}$, respectively, have contributions from interactions (primarily excluded volume), thermal diffusion, and myosin activity, and are proportional to the translational and rotational velocities of the rigid filaments.

To implement this idea in the agent based simulation~\cite{yang2010swarm}, we
need to relate the centre-of-mass velocity of the filament $\bfv_i$ and the rotational velocity $\omega_{i}$ of the corresponding forces and torques acting on the filament. 
Decomposing the centre-of-mass velocity of the filament $i$ as, $\bfv_i = v_{\parallel,i} {\hat \bfn}_i + v_{\perp,i} {\hat \bfm}_i$, where $\hat \bfm_i$ is perpendicular
to $\hat\bfn_i$, we have,
\bea
v_{i,\parallel}(t)&=&\frac{1}{\gamma_\parallel} \bigg( \sum_{j\neq i} {\bf F}_{ij,\parallel} + \xi_{\parallel} {\hat \bfn}_i + {\bf F}_1 + {\bf F}_{2,\parallel} + {\bf F}_{3,\parallel} \bigg)
\nn \\
v_{i,\perp}(t)&=&\frac{1}{\gamma_\perp} \bigg( \sum_{j\neq i} {\bf F}_{ij,\perp} + \xi_{\perp} {\hat \bfm}_i + {\bf F}_{2,\perp} + {\bf F}_{3,\perp} \bigg) \nn
 \\
\omega_{i}(t)&=&\frac{1}{\gamma_r} \bigg( \sum_{j\neq i} {\bf M}_{ij,r} + \xi_r + {\bf M}_2 + {\bf M}_3  \bigg) 
\label{eq:filament}
\eea
where $ {\bf F}_{ij}$ and ${\bf M}_{ij}$ are the passive contributions to the force and torque on the actin filament $i$ from filament $j$, 
$\xi_{\parallel}$, $\xi_{\perp}$ and $\xi_{r}$ are the corresponding contributions from thermal noise, and ${\bf F}_1, \ldots, {\bf M}_3$ are the purely active contributions induced by
myosin-II minifilaments.
The friction coefficients are given by $\gamma_{\parallel}=\gamma_{p}l/10$, $\gamma_{\perp}=2 \gamma_{\parallel}$ and $\gamma_{r}=\gamma_{\parallel}l^2/6$~\cite{doi,howard}.
 All noises are drawn from independent Gaussian distributions with zero mean and variance
 $=2 k_B T \gamma_k /\Delta t$, where 
$k \equiv \parallel, \perp$ or $r$. 
 We use Eq.\,\ref{eq:filament} to update the 
centre-of-mass position and orientation of the $i$-th filament in our Brownian dynamics simulation~\cite{yang2010swarm}.

Since we model actin filaments as a rigid string of beads, the passive contributions to the forces and torques 
on filament $i$
from filament $j$, namely, $ {\bf F}_{ij}$ and ${\bf M}_{ij}$, must be calculated from a bead-bead interaction potential,
\bea
V_{ij}(r) & = &  \sum_{m, m' = 1}^{N_b} 4 \epsilon_a \bigg(\frac{{b}^2}{\vert \bfr_{im} - \bfr_{jm'} \vert^2 + \delta^2}\bigg)^{6} + V_0^{\prime} +V_2^{\prime} \;  \vert \bfr_{im} - \bfr_{jm'} \vert^2  \,\,\,\,\,\,\,\,\, \mbox{for} \,\, \vert\bfr_{im} - \bfr_{jm'}\vert \leq b  \nonumber \\
         & =  &   0    \,\,\,\,\,\,\,\,\,\, \mbox{for} \,\, \vert \bfr_{im} - \bfr_{jm'}\vert > {b}
        \label{eq:pot}
\eea
where $\vert \bfr_{im} - \bfr_{jm'} \vert$ is the distance between the $m$-th bead of $i$-th filament and $m'$-th bead of $j$-th filament
and values of constants $V_{0}^\prime$ and $V_{2}^{\prime}$, are chosen so that the potential and force are continuous at cutoff $r=b$. Here we set the filament bead size $b=1.6\sigma$, and the potential parameters $\epsilon_a=1$ and $\delta=0.8$,
in units of $\epsilon$ and $\sigma$, respectively. With this choice of parameters, an actin filament composed of $N_b=15$ beads, has a mean equilibrium length $l=24\sigma$.

Finally, we describe the form of the active contributions to the forces and torques that arise from the 
interaction of myosin-II minifilaments in $z=2$ with actin filaments in $z=1$. These active forces and torques can only be manifest if some of the myosin heads of the multivalent myosin filament bind and hold on to the
dense Arp 2/3 static meshwork below it (see Fig.\,1). Now instead of explicitly including a structural model for the bulky myosin-II minifilaments, we will simply incorporate the many-body contributions to the active forces and  torques coming from myosin-II minifilaments, using a carefully devised system of extensional and torsional springs, as shown in Fig.\,\ref{fig:rules}.

\begin{enumerate}

\begin{figure}[H]
\centering
\includegraphics[height=0.4\linewidth]{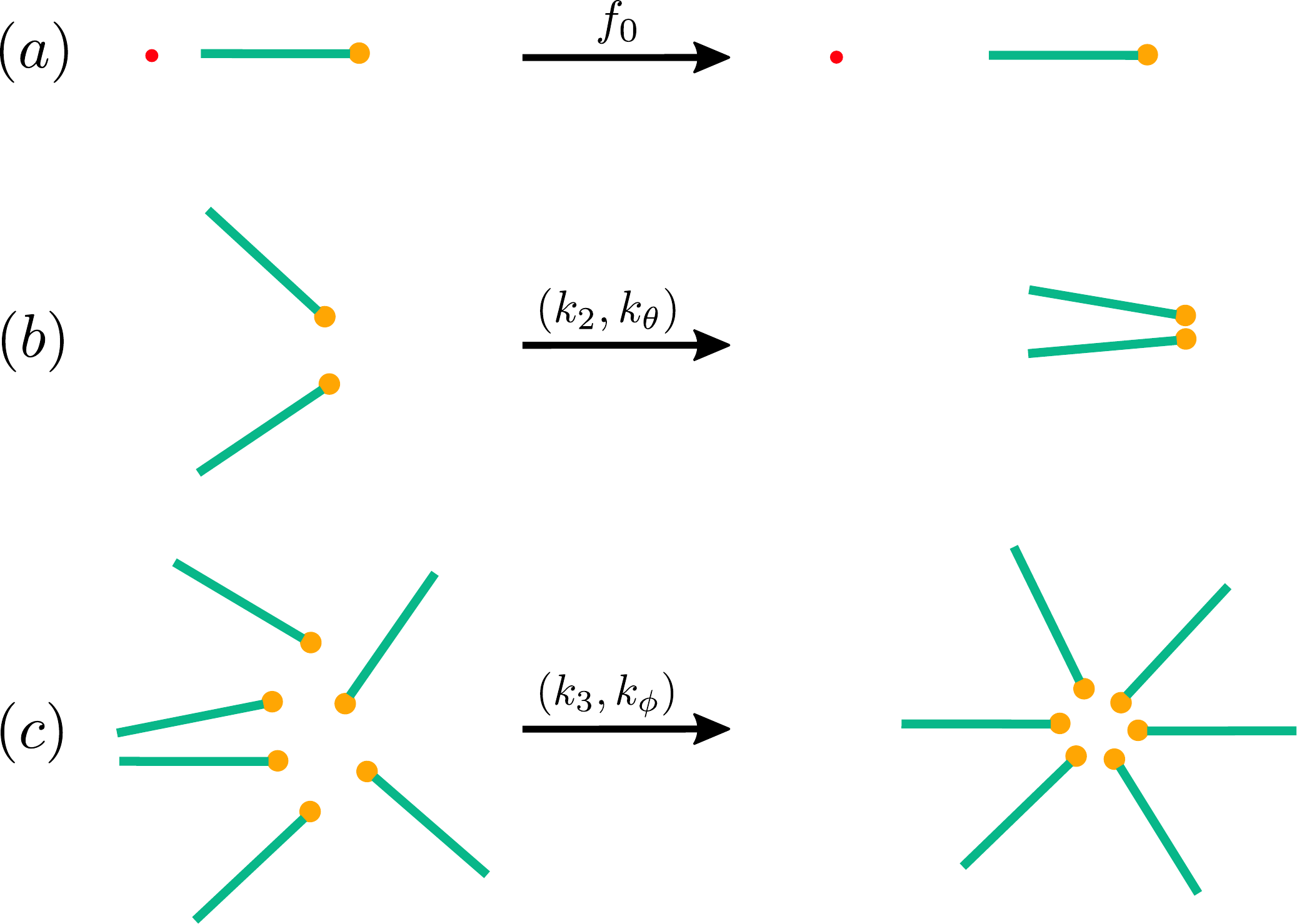}
\caption{Schematic showing the consequences of the active forces and torques induced by Myo-II minifilament on dynamic actin filaments realised in terms of propulsion force and extensional and torsional springs. (a) Single filament contribution: Myosin driven active current $v_0 \hat \bfn$ along the filament orientation pointing to the $+$ end (orange dot),
leading to propulsion away from the reference red dot. (b) Two filament contribution: Myosin driven forces and torques bring together and align two filaments leading to {\it bundling}. This is realised by having an extensional spring ($k_2$) and  torsional spring ($k_\theta$) between the filaments. (c) Multiple filament contribution: Myosin driven forces and torques bring together and reorient multiple filaments leading to the formation of an {\it aster}. This is realised by having extensional springs ($k_3$) and  torsional springs ($k_\phi$) between pairs of filaments.}
\label{fig:rules}
\end{figure}

\item Single filament contribution

Referring to Eq.\,\ref{eq:filament}, 
${\bf F}_1=f_0 \hat \bfn$ is the single filament contribution to the active force driven by a myosin minifilament, where a few myosin heads are attached to the filament and the heads at the other bipolar end of the myosin minifilament are attached to the static actin meshwork.
This leads to an active translation current $\propto v_0 \hat \bfn$, of magnitude $v_0 =f_0/{\gamma_\parallel}$ along its polar orientation $\hat \bfn$ (pointing towards the $+$ end of the filament, see Fig.\,\ref{fig:rules}a). 
In principle, an isolated actin filament 
can also contribute to an active rotational current. We will however assume that the active orientational decorrelation time $\tau_a$ is larger than the rotational diffusion time $\tau_r$
of thermal origin or collision time $\tau_c$ with other filaments. With the filament concentration and filament length under consideration, we have $\tau_c < \tau_r \ll \tau_a$, and thus
 the late-time diffusion coefficient of a single actin filament is set by $D_a=\frac{1}{2}v_0^2\tau_c$.

\item  Two filament contribution 

A myosin minifilament whose heads attach to two dynamic actin filaments at $z=1$ at one end and to the static actin meshwork at the other end,
induces contractile flows that lead to bundling and polar alignment (Fig.\,\ref{fig:rules}b). 
By a polarity sorting mechanism~\cite{koster2016actomyosin,koenderink}, this brings 
the $+$ ends of the dynamic actin filaments together (see Fig.\,\ref{fig:rules}b). This contributes to the 2-filament active forces and torques, ${\bf F}_2$ and ${\bf M}_2$, in Eq.\,\ref{eq:filament}, 
realised here by having an extensional spring (stiffness $k_2$) attached to the $+$ends of the filament pair and a torsional spring (stiffness $k_\theta$), when the filaments are within a cutoff distance (here, taken to be $2 l/3$).
Since the extensional and torsional spring stiffneses have the same molecular origin, they must be related, dimensional considerations suggest $k_2=4 k_\theta/l^2$~\cite{liverpool2003,ahmadi2006}. 
 
\item  Three and multiple filament contribution

Myosin minifilaments whose heads attach to three or more dynamic actin filaments at $z=1$ at one end and to the static actin meshwork at the other end,
induces contractile flows that lead to the formation of bundles and orientational patterns such as asters (Fig.\,\ref{fig:rules}c). This also involves the 
polarity sorting mechanism that brings 
the $+$ ends of the dynamic actin filaments together (Fig.\,\ref{fig:rules}c). This leads to 
${\bf F}_3$ and ${\bf M}_3$, the 3 (or multiple)-filament contributions to the active forces and torques in Eq.\,\ref{eq:filament}.
This is realised by extensional and torsional springs of stiffness $k_3$ and  $k_\phi$, respectively, that operate only when three filaments or more are within a cutoff distance $2 l/3$ of each other. The torsional spring has a rest angle equal to $\pi/n$, where $n$ is the number of actin filaments involved.
As before, dimensional considerations suggest $k_3=4 k_\phi/l^2$~\cite{liverpool2003,ahmadi2006}. 
Note that the salient effects of multivalency of the Myo-II minifilaments appear via these multifilament contributions; thus in our simulations we can turn off the effects of multivalency by simply setting $k_3=k_\phi=0$.
\\

\end{enumerate}

With the forces and torques in place, we numerically integrate the Brownian dynamics equations updating
the position of membrane proteins (following Eq.\,\ref{eq:particle}) and the position and orientation of actin filaments (as given in Eq.\,\ref{eq:filament}) using a Verlet integration scheme~\cite{frenkel} with an integration time step $\Delta t$. For more details on units and parameter values used in the simulation, see Appendix.

{\it Turnover of dynamic actin filaments and myosin}: 
Apart from stratification and multivalency, the other critical feature that we need to include is the turnover of the actomyosin machinery.
The microscopic physics of turnover of myosin minifilaments and dynamic actin is subtle and involves many distinct molecular mechanisms, such
as cooperative unbinding-rebinding of myosin heads and actin filaments, fragmentation of actin filaments by enzymes such as cofilin, depolymerisation, nucleation and recruitment~\cite{guha2005,medeiros2006,haviv2008,salbreux2012,sonal2019,mukhina2007,keren2017}.
Turnover releases the buildup of stresses generated by contractile flows and consequent jamming of 
actin and myosin minifilaments, and aspect that
is necessary for the maintenance of the nonequilibrium steady state.
In our coarse-grained simulation we simply implement turnover by removing actin filaments with a rate $k_r$, modelled as a Poisson process,  
We ensure that in the process, the mean number of actin filaments is held fixed, so that when a filament is removed, we introduce another in a random spatial location with random orientation. 
We have varied $k_r$ over the range $10^{-4} - 10^{-1}$, in simulation units.

Before we end this section, we wish to emphasize the simplicity of our coarse-grained agent based  simulation - it incorporates the minimal features necessary for observing the nonequilibrium steady state, viz., stratification, turnover and 
 multivalency of active force generators, without explicitly including the structural aspects of the bulky myosin-II minifilaments.  It does so by taking into account its effect on {\it currents}, {\it forces} and {\it torques}
applied to single and multiple actin filaments; indeed  we show that 
 abrogation of any one of the three features, leads to a loss of the desired phenotype. This simplicity allows us to  monitor long time dynamics, which would otherwise have been prohibitively difficult. 
 
\section{Orientational patterns of actin and the 
nonequilibrium steady state}
\label{sect:Patterns}

We first explore the orientational patterns displayed by the actin filaments in layer, $z=1$  in the absence of  the membrane proteins at $z=0$ (Fig.\,\ref{fig:schematic}). Recall that we are working in the regime where the overall concentration of the dynamic actin filaments is low, i.e., in the absence of activity, we are in the so-called dilute regime~\cite{doi}. Despite this,  active contractile stresses drive the filaments to form clusters with distinct orientation patterns, as in~\cite{koster2016actomyosin}. 

We define local coarse-grained fields, the actin filament density 
$c(\bfr,t)=\sum_i{\delta(\bfr -\bfr_i(t))}$
and  actin filament polar orientation 
$\bfn(\bfr,t)\,c(\bfr,t)=\sum_i \hat \bfn_i(t)\,\delta(\bfr -\bfr_i(t))$
where $\bfr_i$ and $\hat \bfn_i$ are the centre-of-mass position and polar orientation of the ${\it i}$th filament, respectively.
To characterise the orientational patterning, we first note that the contractile flows lead to strong concentration fluctuations which in turn influence the orientation correlations.
Because of the strong concentration fluctuations, one cannot use a global orientation order parameter to describe the phases observed in the simulations. 
To do so, we first compute the coarse-grained filament concentration profile or the concentration correlation function, from which we extract a correlation length. This defines the spatial scale of a ``cluster''. We compute the net orientation of each cluster, by
projecting the orientation of individual filaments belonging to the cluster onto the mean orientation for that cluster. The polar order parameter $\langle P\rangle$ is then defined as an average over all clusters (and a further average over time and independent realisations).
 
We find that  the orientation patterns within each cluster, can be described as a polar bundle, an aster or a spiral. 
 To characterise spatially varying orientation patterns, we need to 
 compute 
 the divergence and
 curl of the coarse-grained orientation field using the following procedure - (i) choose a coarse graining cell $\Omega(\bfr)$ of linear dimension $2\, l$, around an arbitrary point $\bfr$, (ii) smear the centre of mass of each filament by an exponentially decaying function, an interpolation scheme that allows us to smoothen the vector fields within the cell $\Omega(\bfr)$,
\be
\bfn(k,l)=\f{\sum_{i \in \Omega(\bfr)} \hat \bfn e^{-\vert \bfr - \bfr_i\vert/\lambda}}{\sum_{i \in \Omega(\bfr)} e^{-\vert \bfr - \bfr_i\vert/\lambda}}
\label{eq:smooth}
\ee
where we choose the decay length $\lambda = 3 \sigma$. This allows us to cleanly compute the divergence and the curl of the vector field.

\begin{figure}[H]
\centering
\includegraphics[height=0.40\linewidth]{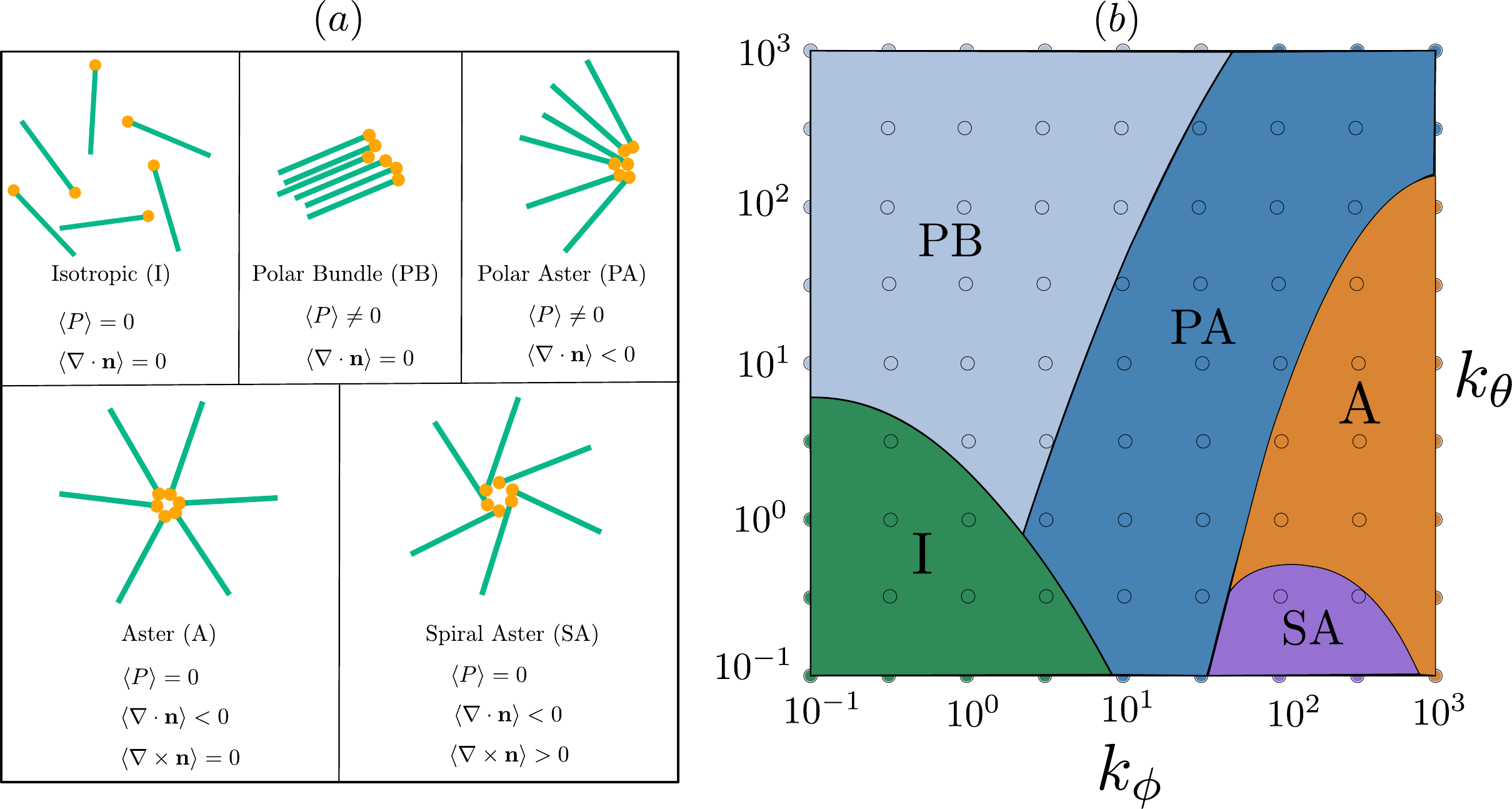}
\caption{Orientational patterns exhibited by polar actin filaments using the Brownian dynamics simulation described in Sect.\,\ref{sect:Simulation}. (a) Characterisation of orientational patterns at steady state - Isotropic (I), Polar Bundle (PB), Polar Aster (PA), Aster (A), and Spiral Aster (SA). (b) Steady state diagram in the space of two-filament and multifilament contributions to the active torque, here realised as torsional springs, $k_\theta$ and $k_\phi$ (see text). 
The circles in the phase diagram depict the points where the simulations were done.
The boundaries between the phases were obtained using a kernel smoothing model with exponential weights.
}
\label{fig:phasedia}
\end{figure}
 Keeping the overall filament concentration fixed, we vary the stiffness
 of the torsional springs ($k_\theta$, $k_\phi$), that in turn alter the relative contributions of the Myo-II induced active torques. As discussed in Sect.\,\ref{sect:Simulation}, since the extensional and torsional springs share a common molecular origin, they are related to each other in a simple manner. From our simulations, we
 identify five distinct orientational phases - isotropic (I),
 polar bundle (PB), polar aster (PA), aster (A) and spiral aster (SA) as shown and characterised in Fig.\,\ref{fig:phasedia}. The order parameters
 characterising these phases are the polar order parameter $\langle P\rangle$, the divergence of the orientation or the splay 
 $\langle \nabla \cdot \bfn\rangle$ and the curl of the orientation or the vorticity 
 $\langle \nabla \times \bfn\rangle$. As shown in Fig.\,\ref{fig:phasedia}, increasing the 
 2-filament contribution to the active torque $k_\theta$ favours polar bundling, while
 increasing the
 multifilament contribution $k_\theta$ favours asters.
 This observation is
 consistent with our earlier theoretical study using a hydrodynamic approach~\cite{gowrishankar2016nonequilibrium,husain2017emergent}. As expected, since these orientation patterns are a consequence of the active forces and torques, the polar bundle and polar asters move along the direction set by their mean orientation and the spiral asters rotate, consistent with the predictions of previous studies ~\cite{kruse,gowrishankar2016nonequilibrium,husain2017emergent}. The patterns described here have been observed in several
 {\it in vitro} reconstitution studies~\cite{backouche2006,murrell2012,koster2016actomyosin,sonal2019}.
 
\begin{figure}[H]
\centering
\includegraphics[height=0.7\linewidth]{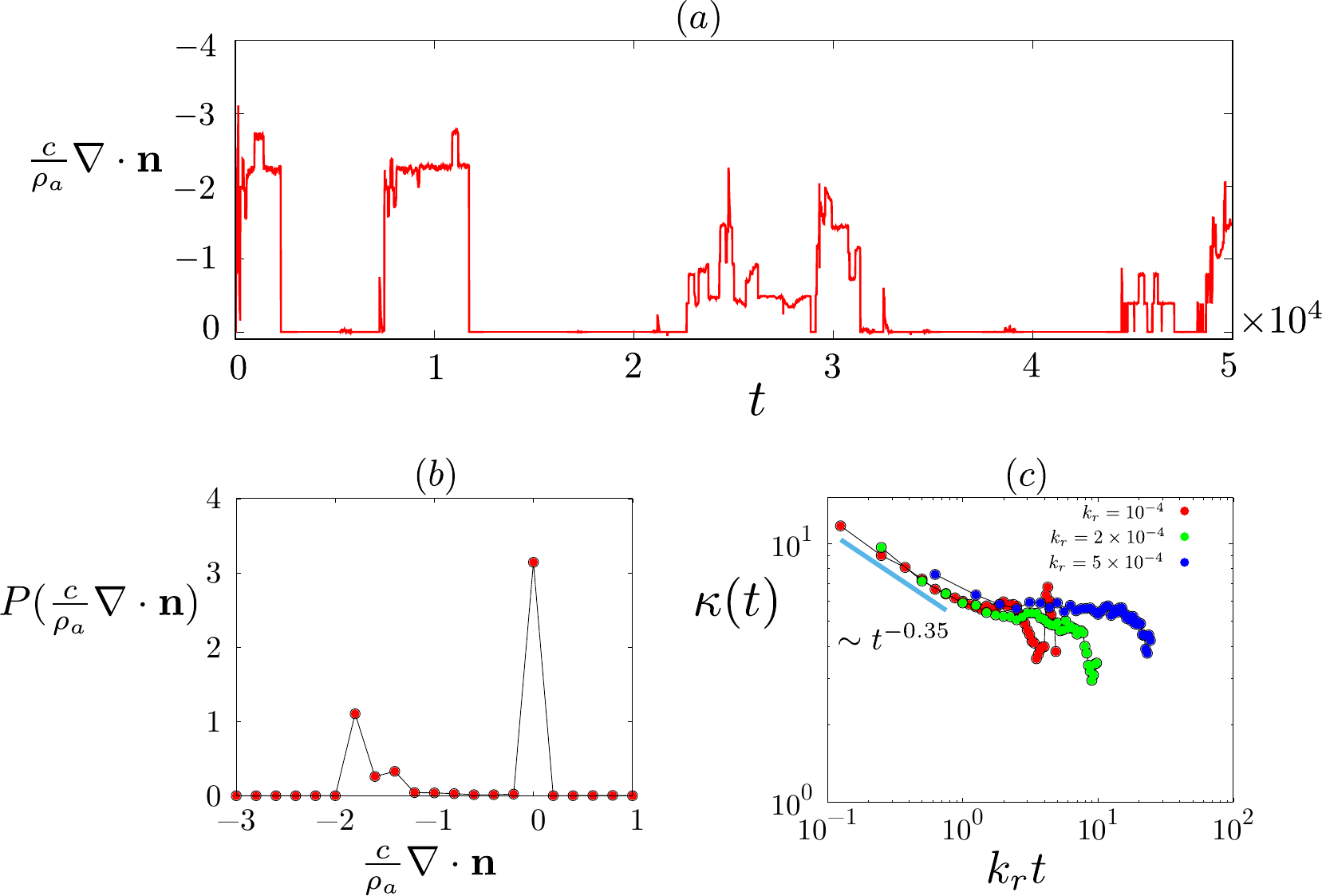}
\caption{Nonequilibrium steady state characterised by intermittent orientational patterns as a result of contractile flows and turnover. (a) For instance, fixing parameters so as to be in the Aster phase of Fig.\,\ref{fig:phasedia}, we monitor the time series of the divergence $\frac{c}{\rho_a}\nabla \cdot {\bf n}$, averaged over an area of size $\frac{L}{4} \times \frac{L}{4}$. This shows large fluctuations ranging from a negative value ($-2$ for one pure aster, where $\bfn = - \hat \bfr$) to $0$.  (b)  The corresponding probability distribution $P(\frac{c}{\rho_a} \nabla \cdot {\bf n})$ measured over $8$ independent initial realisations and $100$  time windows at steady state shows a spread of values from $-2$ to $0$.
Here, torsional spring stiffness $k_\theta=5$, $k_\phi=500$, and turnover rate $k_r=2\times 10^{-4}$. (c) Time dependent kurtosis $\kappa(t)$ of $\frac{c}{\rho_a}\nabla \cdot {\bf n}$ shows a power-law divergence as $k_r t \to 0$, where $k_r$ is the turnover rate. This is characteristic of an intermittent nonequilibrim state. The corresponding $\kappa(t)$ for a Gaussian distributed variable would take a constant value equal to $3$.
}
\label{fig:tover1}
\end{figure}

So far we have included only two ingredients of the active composite, namely stratification and multivalency. In this case, 
once steady state is reached, the orientational patterns described in Fig.\,\ref{fig:phasedia} remain the same, although some will exhibit a mean translation (I, PB, PA) and others (SA) a mean rotation. Now when we introduce the third ingredient, namely steady turnover of the filaments, these patterns undergo dynamic remodelling. In our simulation, we invoke a stress dependent turnover of actin filaments, which relaxes the build up of local contractile active stress
$\sigma_{act}\propto  - c\, (\nabla \cdot {\bfn})$~\cite{salbreux2012,fakhri2018}. This is consistent with recent {\it in vitro} studies~\cite{koster2016actomyosin,sonal2019}, and with observations of turnover of actomyosin in the context of tissue remodelling~\cite{banerjee2017}. The orientation patterns, such as asters, show intermittent fluctuations as seen in  Fig.\,\ref{fig:tover1}(a) - the  time series of  
$\frac{c}{\rho_a}\nabla \cdot {\bf n}$ - where $\rho_a$ is the mean number density of filaments,
and ranges from $-2$ (a pure aster with $\bfn = - \hat \bfr$) to $0$. This is reflected in the skewed probability distribution of the net divergence shown in Fig.\,\ref{fig:tover1}(b). Intermittent dynamics of a statistical variable $X(t)$, here $\frac{c}{\rho_a}\nabla \cdot {\bf n}$, 
identified by alternating periods of quiescence 
and large changes over short times (Fig.\,\ref{fig:tover1}(a)), shows up in the behaviour of the kurtosis, $\kappa(t) = S_4(t)/S_2^2(t)$, the ratio of the fourth central moment and fourth power of the standard deviation of the statistical quantity~\cite{frisch1995,das2016}.
Figure\,\ref{fig:tover1}(c), a plot of the  
kurtosis $\kappa(t)$ versus time, scaled by the turnover rate $k_r$, shows a power-law divergence as $t\to 0$ - a signature
of intermittency characterising the nonequilibrium steady state~\cite{das2016}.
These results are entirely consistent with our earlier hydrodynamic theory~\cite{gowrishankar2012active,gowrishankar2016nonequilibrium,chaudhuri2011spatiotemporal}, and recent {\it in vitro} studies~\cite{koster2016actomyosin,sonal2019}.

\section{Activity enhances multi-particle encounters}

Having described the orientational patterning and nonequilibrium dynamics of actin filaments driven by Myo-II minifilaments, we 
ask how these might affect the dynamics of membrane proteins in the $z=1$ layer of the stratified active composite surface. From
the combined active dynamics of the actin filaments and membrane proteins interacting with actin (Eqs.\,\ref{eq:particle}, \ref{eq:filament}), we see that the contractile flows generated by the active forces and torques on single and multiple actin filaments, draw in the bound membrane proteins. Following this, the membrane proteins may 
unbind from the filaments, resulting locally in high concentrations of free proteins that can engage in multiple binary and multiparticle encounters with each other. This is best seen in the Aster phase of Fig.\,\ref{fig:phasedia}.

\begin{figure}[H]
\centering
\includegraphics[height=0.24\linewidth]{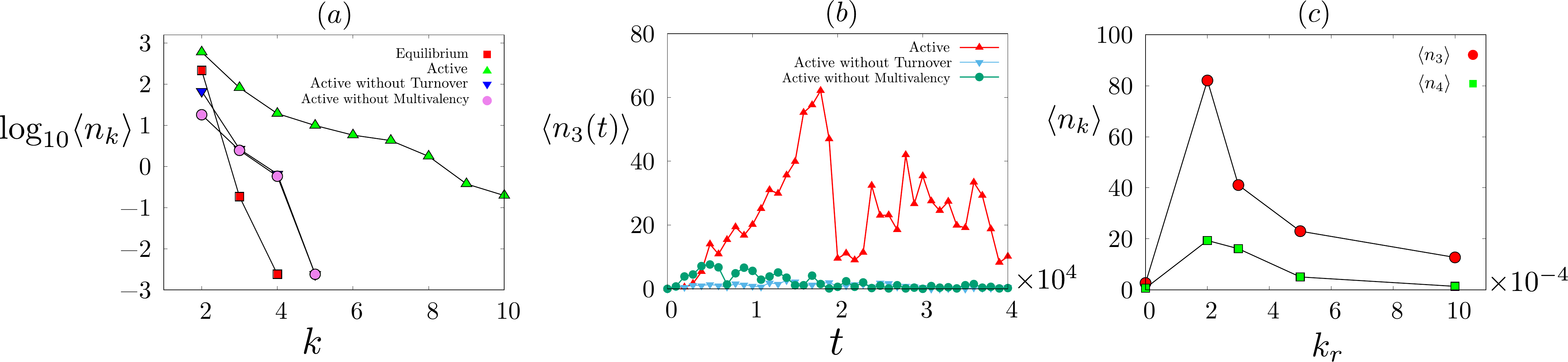}
\caption{Activity and turnover enhance multiparticle encounters. (a) Mean number of $k$-clusters, $\langle n_k \rangle$ versus with cluster size $k$ (y-axis in  $\log_{10}$ scale). Observe the enhancement of the number of large clusters in the active case with both turnover and multivalency (green triangle), compared to the equilibrium limit (red square). Abrogation of turnover (blue triangle) or multivalency (purple circle)
leads to a significant reduction in the number. (b) Time series of  
$\langle n_3(t) \rangle$ shows large intermittent fluctuations  in the nonequilibrium steady state with turnover and multivalency. Fluctuations are dramatically suppressed in the absence of turnover or multivalency. 
(c) Mean number of $k$-clusters $\langle n_k \rangle$ for $k=3$ and $k=4$ shows non-monotonic behaviour with filament turnover rate $k_r$.  }
\label{fig:kmer1}
\end{figure}

To quantify the extent of multi-particle encounters, we compute both the number of clusters of size $k$ and their lifetimes. Figure\,\ref{fig:kmer1}(a)
shows  the average number of clusters of size $k$, denoted as $\langle n_k \rangle$, that appear within a time window (here, we take it to be $10^3$) at steady state. 
(Note that to define a cluster, we need a inter-particle distance cutoff, which we take to be $1.2\,\sigma$). 
In the equilibrium limit, obtained by setting the binding of membrane proteins to the active actin filaments to zero, we see that thermal motion alone leads to transient clusters that are predominantly dimers ($k=2$)
and rarely  trimers or higher $k$-mers ($k\geq 3$).
This is in striking contrast to the active steady state with turnover, which shows a significant fraction of large clusters ($k$ up to $10$, as seen in Fig.\,\ref{fig:kmer1}(a)).

We emphasize that these multiparticle encounters are realised only in the active nonequilibrium steady state, which is contingent on
stratification, turnover and multivalency of force generators. Significantly, when we abrogate either turnover ($k_r=0$) or multivalency ($k_3=k_\phi=0$), we loose the larger clusters. 

In the nonequilibrium steady state, the number of $k-$clusters shows the same intermittent dynamics as mean aster density.
Figure\,\ref{fig:kmer1}(b) shows the time series of the number of $3$-clusters $\langle n_3(t) \rangle$ 
in the nonequilibrium steady state. The large temporal fluctuations
of  $\langle n_3(t) \rangle$ are suppressed in the absence of turnover and multivalency of the active force generators. 
Interestingly, $\langle n_k \rangle$ shows a non-monotonic dependence on the  turnover rate $k_r$ (Fig.\,\ref{fig:kmer1}(c)), a prediction that can be tested in reconstitution experiments, such as~\cite{koster2016actomyosin}.

To summarize, the 
formation of large and intermittent clusters in the nonequilibrium steady state is contingent on stratification, turnover and multivalency,
in agreement with experiments on reconstituted actomyosin on a supported bilayer~\cite{koster2016actomyosin}.

\section{Activity influences chemical reaction kinetics at the cell surface : reaction optimum}

Since many signalling molecules at the cell surface interact with cortical actin, it is reasonable that most biochemical reactions on the cell membrane are strongly influenced by the actomyosin cortex. 
For instance, there have been proposals that trapping of membrane proteins by the  cortical actin mesh underlying the cell membrane can  lead to  enhancement of reaction kinetics~\cite{haugh,
voituriez2008,benichou2011,kalay2012reaction,kalay2012confinig,
wolde2005,johnson}. Since in these models there is no dynamical feedback between the chemical reactants and the cortical mesh, there is no mechanism by which spatiotemporal control and regulation can be effected.  
 Our observation that multi-particle encounters of membrane proteins are more frequent when they are driven by active forces and torques, together with turnover of the active machinery, 
 suggests an additional mechanism for enhancement of 
 chemical reaction kinetics along with its spatiotemporal regulation~\cite{chaudhuri2011spatiotemporal}.

 To fix ideas, let us consider 
  a reversible bimolecular reaction between two membrane protein species $A$ and $B$ reacting to form a complex $C$,
$ A + B \underset{k_{b}}{\stackrel{k_{f}}{\rightleftharpoons}} C$,
where $k_{f}$ and $k_{b}$ are the effective forward and backward reaction rates, respectively. For the forward reaction to be realised, the A and B proteins need to first diffuse towards each other, close enough 
(we take the scale of this reaction zone to be $1.2\,\sigma$), so as to engage in a chemical bonding, and then associate to form $C$ with a rate $k_{a}$ modelled as a Poisson process. The effective
backward reaction, involves a chemical dissociation of the complex $C$ with a rate $k_{d}$ modelled as a Poisson process and a subsequent escape of the products $A$ and $B$ from the reaction zone.
Note that the individual chemical reactions are equilibrium processes, the only role that activity plays in these chemical reactions is in creating situations where the local reactant density is high.

\begin{figure}[H]
\centering
\includegraphics[height=0.70\linewidth]{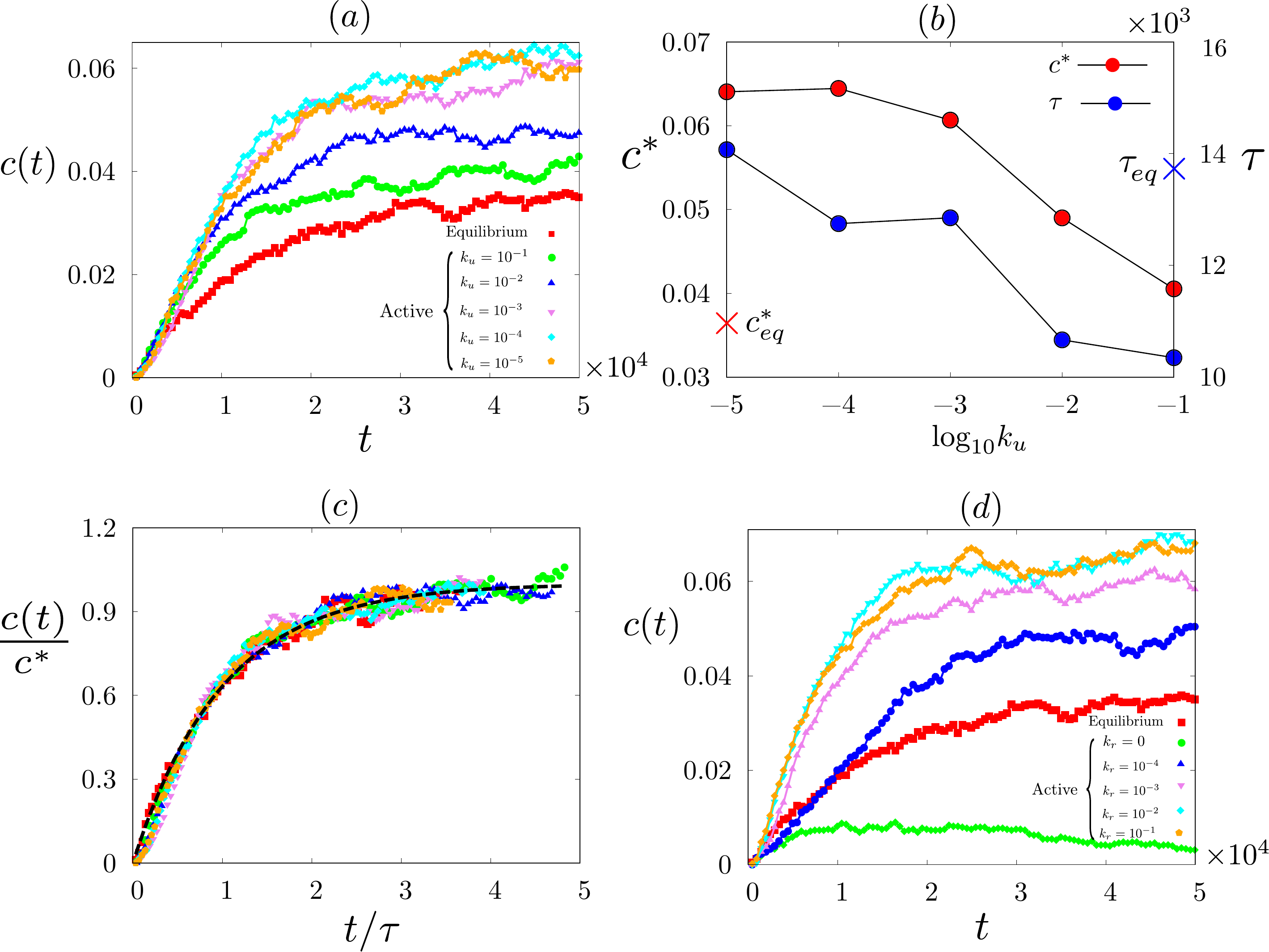}
\caption{(a) The fraction of C proteins $c(t)$ increases with time as the bimolecular reaction proceeds; in the nonequilibrium steady state with contractile flows and turnover, both the reaction yield and net reaction speed, increase with increasing actin binding affinity (or decreasing unbinding rate $k_u$), keeping the turnover rate at $k_r=5\times 10^{-4}$. 
(b) Reaction yield $c^{\ast}$, obtained from the saturation concentration, and the net reaction time $\tau$, obtained from the initial slope of the growth curve, decrease with unbinding rate. For reference, we display the value of $c^{\ast}$ and $\tau$ for the equilibrium reaction. 
 (c) Using the extracted values of $c^{\ast}$ and $\tau$, we display a scaling collapse of all the data with different values of $k_u$.
The dashed line in (c) is the scaling curve $(1-e^{-\frac{t}{\tau}})$. 
(d) The fraction of C proteins $c(t)$ in the nonequilibrium steady state versus time for different turnover rates $k_r$ keeping the protein unbinding rate fixed $k_u=10^{-4}$. 
The reaction yield $c^{\ast}$ increases with increasing turnover rate $k_r$, and is higher than the equilibrium reaction.
One expects the reaction yield to start {\it decreasing} when $k_r \gg k_u$. 
Note that without turnover ($k_r=0$),  the reaction yield is extremely small.  Throughout this panel, we have fixed the torsional spring stiffness $k_\theta=5$, $k_\phi=500$, binding rate $k_b=0.5$, association rate $k_a=4\times10^{-3}$ and disassociation rate $k_d=10^{-4}$.}
\label{fig:ck}
\end{figure}

The fraction of C proteins $c(t) =2N_C(t)/N$, starting from an equal number of $A$ and $B$ proteins, appears to
grows as $c(t) = c^{\ast}\l 1-e^{-\frac{t}{\tau}}\r$,  even 
in the nonequilibrium steady state, simulated for different values of $k_u$ and  turnover rate $k_r>0$ (see, Fig.\,\ref{fig:ck}(a-d)). This form is exactly what one would expect for a mass action bimolecular reaction,
\be
 \frac{d\rho_C(t)}{dt} = k_{f} \; \rho_A(t) \rho_B(t) -  k_{b} \; \rho_C(t)
\label{eq:reaction}
\ee

where, $\rho_A(t)$,  $\rho_B(t)$ and  $\rho_C(t)$ are the number density of $A$, $B$ and $C$ molecules, respectively.
This is reinforced by constructing a scaling plot over a range of values of $k_u$ (Fig.\,\ref{fig:ck}(c)) using the extracted 
parametric dependence of $c^{\ast}$ and ${\tau}$ with $k_u$ (Fig.\,\ref{fig:ck}(b)).

From Fig.\,\ref{fig:ck}(a,b), we see that the reaction yield $c^{\ast}$ is significantly larger in the nonequilibrium steady state with active driving and turnover compared to its value at equilibrium and shows a strong dependence on actin binding affinity $k_u$ of the reactants. Further, the reaction proceeds faster as seen in the plot of the reaction time $\tau$ (Fig.\,\ref{fig:ck}(b)). 

In general, this is due to a combination of faster advective transport and enhancement in the local concentration of reactants as a consequence of a larger binding affinity to actin and the active contractile flows. In Fig.\,\ref{fig:ck}, since the values of the reactin parameters are such that $k_a/D < 0.05$, the chemical reactions are largely association rate limited~\cite{agmon1990theory}, and so the dominant mechanism for .

Finally, we see that fixing the binding affinity, the reaction yield increases with increasing turnover rate. This is
due to the combination of stochastic active contractile flows and turnover in the nonequilibrium steady state, and can be thought of as contributing to both an enhanced (active) temperature and an increase in local concentration of reactants. Note that when the turnover is switched off and the actin asters are static, there is very little increase in the  of the reaction yield Fig.\,\ref{fig:ck}(d).

\section{Discussion}

In this paper we have developed a new coarse grained agent based simulation to describe the dynamics of a stratified active composite of a membrane, comprising lipids and proteins,  juxtaposed with an active
actomyosin layer.
Recent {\it in vitro} studies and computer simulations have shown that in order
 to {\it maintain} a nonequilibrium steady state characterised by steady active contractile flows, dynamic orientational patterning and dynamic protein clustering, one 
needs in addition to the stratification of the active machinery, a constant {\it turnover} of active components, that give rise to a 
dynamic force patterning.
Here we bring together, in a rather simple coarse-grained agent based simulation, all the  ingredients necessary for observing the nonequilibrium steady state, viz., stratification, turnover and 
 multivalency of force generators. 
 Stratification relieves us of the obligation to include 
 steric hindrance between molecular components that reside in different strata, though it is important to retain steric interactions between molecules that reside in the same 
 layer. Even so such a computation would have been a daunting
 task, since including a representation of the essential structural features of Myo-II minifilaments in the cortical layer, would have made it exceedingly difficult to monitor long time dynamics. Thus, our coarse grained simulation method has intrinsic value, since it takes into account the effects of myosin minifilaments on currents, {\it forces} and {\it torques} applied to single and multiple actin filaments, without explicitly incorporating its structural details.
 Consistent with experimental observation, we 
 we show that inhibition of any one of the three ingredients, leads to a loss of the desired phenotype.

We show that this simple coarse grained model, that incorporates the fluctuating active forces and torques in a stratified geometry, recapitulates the nonequilibrium steady states observed both  {\it in vivo}~\cite{rao2014active,goswami2008nanoclusters,gowrishankar2012active,guha2005,medeiros2006,reichl2008,
yang2012} and in properly designed reconstitution experiments~\cite{koster2016actomyosin,sonal2019,fakhri2018,amitdas2020}. By coupling this actomyosin dynamics to the dynamics of molecules on the two dimensional membrane bilayer, we find that  this automatically drives {\it multiparticle} encounters and dynamical clustering.
This provides the motivation to study the influence of active stresses on the dynamics of generic bimolecular chemical reactions, $A + B\,\,{\rightleftharpoons}\,\,C$ on the cell surface~\cite{chaudhuri2011spatiotemporal}. 

We conclude this section, by pointing to the simplicity of the coarse grained agent based  simulation, in that it incorporates all the  ingredients necessary for observing the nonequilibrium steady state
viz., stratification, turnover and 
 multivalency of force generators, without having to explicitly include the structural aspects of myosin minifilaments.  It does so by taking into account its effect on {\it currents}, {\it forces} and {\it torques}
applied to single and multiple actin filaments; indeed  we show that 
 inhibition of any one of the three ingredients, leads to a loss of the desired phenotype. This allows us to  monitor long time dynamics, which would otherwise have been exceedingly difficult.

The most relevant consequence of our work is that it is a simple model that can be easily extended to address a variety of relevant physical situations, such as 
an extension to multicomponent bilayer capable of phase segregation~\cite{saha2021}.

\section{Acknowledgement}
We thank S. Mayor, J. P. Banerjee, S. Talluri, S. Das and A. Kumar for useful discussions. SKRH thanks the Simons Centre at NCBS for computational facilities and hospitality. RM acknowledges funding from the European Union’s Horizon 2020 research and innovation programme under Marie Sklodowska-Curie grant agreement No. 893128. MR acknowledges support
from a JC Bose Fellowship. 

\appendix

\section{Parameters values used in simulation} 

Throughout our simulation, the units of length, time and energy are set by $\sigma$, $\gamma_p$ and $\epsilon$ (Table\,\ref{table:Units}).
All results presented here are for $N_p=300$ membrane proteins and $N_a = 64$ actin filaments in a two dimensional (2d) area of linear dimension $L =400$ with periodic boundary conditions (PBC). In our Brownian dynamics simulations runs, the time update is $\Delta t \sim 2 \times 10^{-3}$, with 
total run time being $t = 5 \times 10^4$. Our initial conditions are chosen from a thermal distribution at temperature $T=1.0$, and all results presented here are averaged over $32$ such independent initial realisations. We have fixed the number of membrane proteins to be $N_p=300$, and number of actin filaments to be $N_a=64$, which gives a number density of $\rho_p=1.8 \times 10^{-3}$ and 
$\rho_a=4\times10^{-4}$, respectively.

All other parameters expressed in natural units are listed in Table.\,\ref{table:Param} for convenience.

\begin {table}[H]
\caption {Natural units - simulation units (S.U.) and real units (R.U.)}
\begin{center}
 \begin{tabular}{||c ||c ||c ||c ||} 
 \hline
Natural Units & Symbol\,[Dimension]       &  S.U. &    R.U.                      \\ [0.5ex] 
 \hline\hline
 
Length (Membrane protein diameter)   &   $\sigma$     $[l]$ &       1  &        $ 10$   $nm$      \\ 
 \hline 

Energy (Inter-protein interaction)  &        $\epsilon$   &             1  &      $4.11 \times 10^{-21}$ $J$         \\ 
 \hline

Membrane protein friction coefficient  &       $\gamma_p$  &   10   &       $ 0.8$  $ pN \mu m^{-1} s $         \\ 
 \hline
 
 Time   &        $ t=l^2 \gamma_p/10\epsilon$ &         1  &       $ 2 \times 10^{-3}$ $s$        \\ 
 \hline
\end{tabular}
 \label{table:Units}
\end{center}
\end{table}

 \begin {table}[H]
\caption {Other parameters expressed in natural units and their ranges}
\begin{center}
 \begin{tabular}{||c ||c ||c ||} 
 \hline
Parameters & Symbol\,[Dimension]       &  Value/Range                     \\ [0.5ex] 
 \hline\hline
 
Membrane protein diffusion constant &  
$D$ [$l^2t^{-1}$] &   $0.1$   \\ 
 \hline
   
Physiological Temperature &         
$T$ [$\epsilon$]  &              $1$     \\ 
 \hline
  
Actin filament bead diameter &    
   $\sigma_b$ [$l$]     & 1.6      \\ 
 \hline 
  
Actin filament length & 
 $l$ [$l$]    &   $24$    \\ 
 \hline
 
Single actin filament propulsion velocity & 
$v_0$ [$lt^{-1}$]    &   $0.2$  \\ 
 \hline
 
Torsional spring stiffness & 
$k_\theta, k_\phi$ [$\epsilon$]  & $10^{-1}-10^{3}$      \\ 
 \hline
 
Extensional spring stiffness  & 
$k_2, k_3$ [$\epsilon l^{-2}$]  & $10^{-1}-10^{3}$ \\ 
\hline

Turnover rate  of actin filaments & 
  $k_r$ [$t^{-1}$]  &   $10^{-1}-10^{-4}$  \\ 
 \hline

 Binding rate of protein to actin filament & 
$k_b$ [$l^2t^{-1}$]    &   $0.5$    \\ 
 \hline
 
Unbinding rate of protein from actin filament & 
 $k_u$ [$t^{-1}$]     &   $10^{-1}-10^{-5}$   \\ 
 \hline
 
Association rate of bimolecular reaction  &  
$k_a$ [$l^2t^{-1}$]   &   $0.004$  \\ 
 \hline
 
Disassociation rate of bimolecular reaction  & 
$k_d$ [$t^{-1}$]    &   $10^{-4}$    \\ 
 \hline

\end{tabular}
 \label{table:Param}
\end{center}
\end{table}

{}

\end{document}